\documentclass[multphys,vecphys]{svmult}

\usepackage{makeidx}     
\usepackage{graphicx}    
\usepackage{multicol}    

\makeindex             

\begin{document}

\title*{The Standard Candle Method for Type II Supernovae and the Hubble Constant}
\titlerunning{Distances from Type II Supernovae}
\author{Mario Hamuy}
\institute{The Observatories of the Carnegie Institution of Washington
\texttt{mhamuy@ociw.edu} }
\maketitle

The ``standard candle method'' for Type II plateau supernovae
produces a Hubble diagram with a dispersion of 0.3 mag, which
implies that this technique can produce distances with a
precision of 15\%. Using four nearby supernovae
with Cepheid distances I find $H_0(V)$=75$\pm$7, and $H_0(I)$=65$\pm$12.

\section{Introduction}
\label{intro}

Type II supernovae are exploding stars characterized by
strong hydrogen spectral lines and their proximity to star forming
regions, presumably resulting from the gravitational collapse of
the cores of massive stars ($M_{ZAMS}$$>$8 $M_\odot$).
These objects display great variations in their spectra and lightcurves
depending on the properties of their progenitors at the time
of core collapse and the density of the medium in which they explode \cite{hamuy03a}.
The plateau subclass (SNe~IIP) constitutes a well-defined family which can
be distinguished by
1) a characteristic ``plateau'' lightcurve \cite{barbon79},
2) Balmer lines exhibiting broad P-Cygni profiles, and
3) low radio emission \cite{weiler02}. These SNe are thought
to have red supergiant progenitors that do not experience significant mass loss
and are able to retain most of their H-rich envelopes before explosion.

Although SNe~IIP display a wide range in luminosity, rendering their use as
standard candles difficult, Hamuy \& Pinto (2002) \cite{hamuy02} (HP02)
used a sample of 17 SNe~II to show that the relative luminosities of these
objects can be standardized from a spectroscopic measurement of the SN ejecta velocity. 
Recently, I confirmed the luminosity-velocity relation \cite{hamuy03b} (H03)
from a sample of 24 SNe~IIP. This study showed
that the ``standard candle method'' (SCM) yields a Hubble diagram with a dispersion of 0.3 mag,
which implies that SNe~IIP can be used to derive extragalactic distances with a precision of 15\%. 
Since the work of H03, Cepheid distances to two SNe~IIP have been published,
bringing to four the total number of SNe~IIP with Cepheid distances.
In this paper I use these four objects to improve the calibration of the
Hubble diagram, and solve for the value of the Hubble constant.

\section{The Luminosity-Velocity Relation}
\label{lv_sec}

\begin{figure}
\centering
\includegraphics[height=75mm]{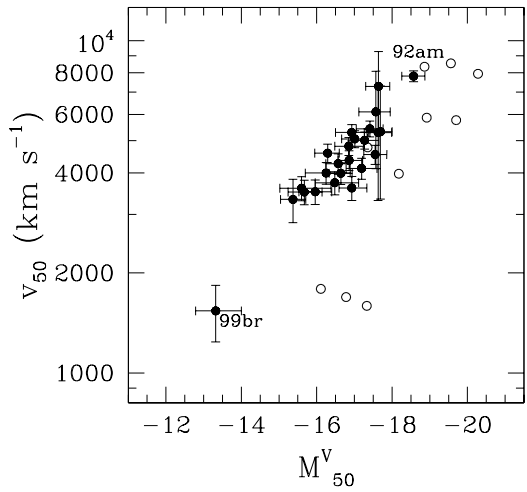}
\caption{Envelope velocity versus absolute plateau $V$ magnitude for
24 SNe~IIP, both measured in the middle of the plateau (day 50)
(filled circles). The expansion velocities were obtained from
the minimum of the Fe II $\lambda$5169 lines. The absolute magnitudes
were derived from redshift-based distances and observed magnitudes
corrected for dust extinction. Open circles correspond to 
explosion models computed by \cite{litvinova83} and \cite{litvinova85}
for stars with $M_{ZAMS}$ $\geq$ 8 $M_\odot$. }
\label{L_v.fig}
\end{figure}

The SCM is based on the luminosity-velocity relation, which permits one to
standardize the relative luminosities of SNe~IIP. Figure \ref{L_v.fig}
shows the latest version, based on 24 genuine SNe~IIP.
This plot reveals the well-known fact that SNe~IIP encompass a
wide range ($\sim$5 mag) in luminosities. This correlation reflects
the fact that while the explosion energy increases, so do the kinetic
energy and internal energies. Also plotted in this figure with open circles
are the explosion models computed by \cite{litvinova83} and \cite{litvinova85}
for stars with $M_{ZAMS}$ $\geq$ 8 $M_\odot$, which reveals
a reasonable agreement with observations.

\section{The Hubble Diagram}

In a uniform and isotropic Universe we expect locally a linear relation
between distance and redshift. A perfect standard candle 
should describe a straight line in the magnitude-log($z$)
Hubble diagram, so the observed scatter is a measure of how standard
the candle is. Next I assess the performance of the SCM
based on the Hubble diagram constructed with the magnitudes and redshifts
given in Table \ref{SN.tab} for 24 SNe.

\begin{table}[ht]
\centering
\caption{Redshifts, Extinction, Magnitudes, and Ejecta Velocities of the 24 Type II Supernovae.}
\begin{tabular}{lcccccc}
\hline 
SN & $v_{CMB}$             & $A_{GAL}(V)$ & $A_{host}(V)$ &   $V_{50}$ & $I_{50}$      & $v_{50}$      \\
   &   (km s$^{-1}$)       &              &               &            &               & (km s$^{-1}$) \\
   &    $\pm$187           & $\pm$0.06    & $\pm$0.3      &            &               &               \\
\hline
1968L   &  321 & 0.219  & 0.00 & 12.03(08) &  ...      & 4020(300)   \\
1969L   &  784 & 0.205  & 0.00 & 13.35(06) &  ...      & 4841(300)   \\
1970G   &  580 & 0.028  & 0.00 & 12.10(15) &  ...      & 5041(300)   \\
1973R   &  808 & 0.107  & 1.40 & 14.56(05) &  ...      & 5092(300)   \\
1986I   & 1333 & 0.129  & 0.20 & 14.55(20) & 14.05(09) & 3623(300)   \\
1986L   & 1466 & 0.099  & 0.30 & 14.57(05) &  ...      & 4150(300)   \\
1988A   & 1332 & 0.136  & 0.00 & 15.00(05) &  ...      & 4613(300)   \\
1989L   & 1332 & 0.123  & 0.15 & 15.47(05) & 14.54(05) & 3529(300)   \\
1990E   & 1426 & 0.082  & 1.45 & 15.90(20) & 14.56(20) & 5324(300)   \\
1990K   & 1818 & 0.047  & 0.20 & 14.50(20) & 13.90(05) & 6142(2000)  \\
1991al  & 4484 & 0.168  & 0.00 & 16.62(05) & 16.16(05) & 7330(2000)  \\
1991G   & 1152 & 0.065  & 0.00 & 15.53(07) & 15.05(09) & 3347(500)   \\
1992H   & 2305 & 0.054  & 0.00 & 14.99(04) &  ...      & 5463(300)   \\
1992af  & 5438 & 0.171  & 0.00 & 17.06(20) & 16.56(20) & 5322(2000)  \\
1992am  &14009 & 0.164  & 0.28 & 18.44(05) & 17.99(05) & 7868(300)   \\
1992ba  & 1192 & 0.193  & 0.00 & 15.43(05) & 14.76(05) & 3523(300)   \\
1993A   & 8933 & 0.572  & 0.05 & 19.64(05) & 18.89(05) & 4290(300)   \\
1993S   & 9649 & 0.054  & 0.70 & 18.96(05) & 18.25(05) & 4569(300)   \\
1999br  &  848 & 0.078  & 0.65 & 17.58(05) & 16.71(05) & 1545(300)   \\
1999ca  & 3105 & 0.361  & 0.68 & 16.65(05) & 15.77(05) & 5353(2000)  \\
1999cr  & 6376 & 0.324  & 0.00 & 18.33(05) & 17.63(05) & 4389(300)   \\
1999eg  & 6494 & 0.388  & 0.00 & 18.65(05) & 17.94(05) & 4012(300)   \\
1999em  &  838 & 0.130  & 0.18 & 13.98(05) & 13.35(05) & 3757(300)   \\
1999gi  &  706 & 0.055  & 0.68 & 14.91(05) & 13.98(05) & 3617(300)   \\
\hline
\end{tabular}
\label{SN.tab}
\end{table}

The CMB redshifts of the SN host galaxies were derived from the observed
heliocentric redshifts. For the 16 SNe with $cz$$<$3000 km~s$^{-1}$
I corrected the redshifts for the peculiar motion of the SN hosts using
the parametric model for peculiar flows of \cite{tonry00} (see H03 for details).
In all cases I assigned an uncertainty of $\pm$187 km~s$^{-1}$, which  corresponds
to the cosmic thermal velocity yielded by the parametric model. 

A convenient choice for SNe~IIP is to use magnitudes
in the middle of the plateau, so I interpolated the observed $V$ and $I$ fluxes
to the time corresponding to 50 days after explosion. In order to use SNe~IIP
as standardized candles it proves necessary to correct the observed fluxes
for dust absorption. The determination of Galactic extinction is under good control
thanks to the IR dust maps of \cite{schlegel98}, which permit one to
estimate $A_{GAL}(V)$ to $\pm$0.06 mag. The determination of absorption in
the host galaxy, on the other hand, is difficult. In H03 I described
a method which assumes that SNe~IIP should all reach the same color toward the end
of the plateau phase. The underlying assumption is that the opacity in SNe~IIP is
dominated by e$^-$ scattering, so they should all reach the temperature of
hydrogen recombination as they evolve \cite{eastman96}.
The method is not fully satisfactory since some discrepancies 
were obtained from $B-V$ and $V-I$ (probably caused by metallicity variations
from SN to SN). An uncertainty of $\pm$0.3 mag can be assigned to this technique
based on the reddening difference yielded by both colors.

The ejecta velocities come from the minimum of the Fe II $\lambda$5169
lines interpolated to day 50, which is good to $\pm$300 km~s$^{-1}$ \cite{hamuy01}.
In the four cases where I had to extrapolate velocities
I adopted an uncertainty of $\pm$2000 km~s$^{-1}$.

\begin{figure}[ht]
\centering
\includegraphics[height=75mm,width=75mm]{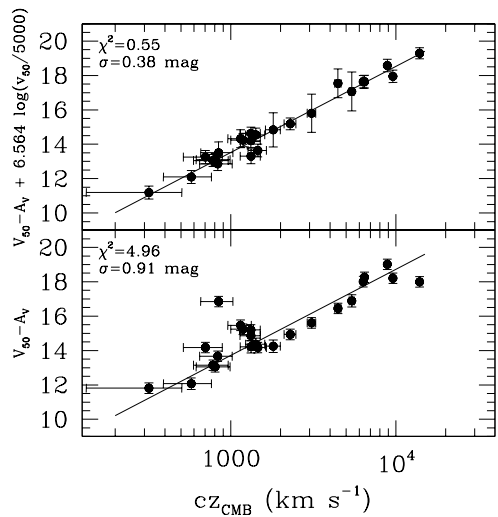}
\caption{(bottom) Raw Hubble diagram from SNe~II plateau $V$ magnitudes.
(top) Hubble diagram from $V$ magnitudes corrected for envelope expansion
velocities. }
\label{hd3.fig}
\end{figure}

The bottom panel of Fig. \ref{hd3.fig} shows the Hubble diagram in the $V$ band,
after correcting the apparent magnitudes for the reddening values, while the top
panel shows the same magnitudes after correction for expansion velocities.
A least-squares fit to the data in the top panel yields the following solution,

\begin{equation}
V_{50} - A_{V} + 6.564(\pm0.88)~log (v_{50}/5000) = 5~log(cz) - 1.478(\pm0.11).
\label{veqn_1}
\end{equation}

\noindent The scatter drops from 0.91 mag to 0.38 mag, thus demonstrating that
the correction for ejecta velocities standardizes the luminosities of SNe~IIP
significantly. It is interesting to note that part of the spread comes from
the nearby SNe which are potentially more affected by peculiar motions of their
host galaxies. When the sample is restricted to the eight objects with
$cz$$>$3,000 km s$^{-1}$, the scatter drops to only 0.33 mag.
The corresponding fit for the restricted sample is,

\begin{equation}
V_{50} - A_{V} + 6.249(\pm1.35)~log (v_{50}/5000) = 5~log(cz) - 1.464(\pm0.15).
\label{veqn_2}
\end{equation}

\begin{figure}[ht]
\centering
\includegraphics[height=75mm,width=75mm]{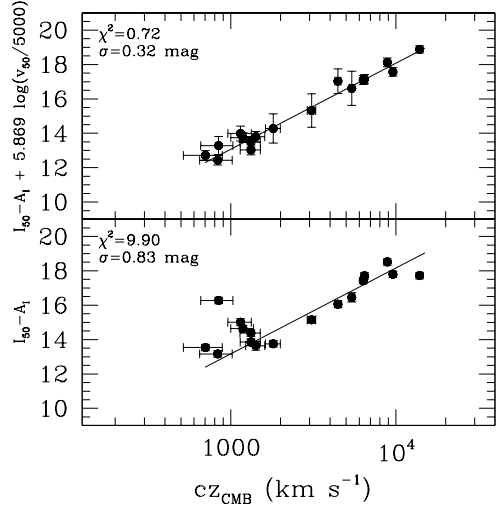}
\caption{(bottom) Raw Hubble diagram from SNe~II plateau $I$ magnitudes.
(top) Hubble diagram from $I$ magnitudes corrected for envelope expansion
velocities. }
\label{hd4.fig}
\end{figure}

Figure \ref{hd4.fig} shows the same analysis but in the $I$ band. In this case the
scatter in the raw Hubble diagram is 0.83 mag, which drops to 0.32 mag
after correction for ejecta velocities. This is even smaller that the 0.38
spread in the $V$ band, possibly due to the fact that the effects of dust
extinction are smaller at these wavelengths. The least-squares fit yields the
following solution,

\begin{equation}
I_{50} - A_{I} + 5.869(\pm0.68)~log (v_{50}/5000) = 5~log(cz) - 1.926(\pm0.09).
\label{ieqn_1}
\end{equation}

When the eight most distant objects are employed the spread is 0.29 mag,
similar to that obtained from the $V$ magnitudes and  the same sample, and the solution is,

\begin{equation}
I_{50} - A_{I} + 5.445(\pm0.91)~log (v_{50}/5000) = 5~log(cz) - 1.923(\pm0.11).
\label{ieqn_2}
\end{equation}

\section{The Value of the Hubble Constant}

The SCM can be used to solve for the Hubble constant, provided the
distance to a nearby SN is known.  If the distance $D$ of the calibrator
is known, and the distant sample is adopted, the Hubble constant is given by

\begin{equation}
H_0(V) = \frac {10^{V_{50} - A_V + 6.249 log(v_{50}/5000) + 1.464}}{D},
\label{h0_v}
\end{equation}

\begin{equation}
H_0(I) = \frac {10^{I_{50} - A_I + 5.445 log(v_{50}/5000) + 1.923}}{D}.
\label{h0_i}
\end{equation}

Among the objects of our sample SN~1968L, SN~1970G, SN~1973R, and SN~1999em have
precise Cepheid distances.  The distances and the corresponding $H_0$ values are
summarized in Table \ref{H0.tab}. SN~1999em is the only object that provides
independent values from the $V$ and $I$ bands, and the results agree remarkably well. 
Within the uncertainties the values derived from the $V$-band magnitudes are in
good agreement for all four objects, and the average proves to be $H_0(V)$=75$\pm$7 km~s$^{-1}$~Mpc$^{-1}$. 

Currently, the most precise extragalactic distance indicators are the peak luminosities
of SNe~Ia. While the HST Key Project yielded a value of $H_0$=71$\pm$2 \cite{freedman01},
Sandage and collaborators derived $H_0$=59$\pm$6 \cite{parodi00}. The difference
is mostly due to systematic uncertainties in the Cepheid distances of the calibrating
SNe. Since SCM is mainly calibrated with Cepheid distances of the HST Key Project,
I conclude that both SNe~Ia and SNe~IIP give consistent results, which lends further
credibility to the SCM.

\begin{table}[ht]
\centering
\caption{The Hubble Constant.}
\begin{tabular}{lcccc}
\hline 
SN & Distance         & Reference         &  $H_0(V)$                 & $H_0(I)$  \\
   & Modulus          &                   &  (km~s$^{-1}$~Mpc$^{-1}$) & (km~s$^{-1}$~Mpc$^{-1}$)           \\
\hline
1968L   & 28.25(15)   & \cite{thim03}     & 77$\pm$15                 & ...       \\
1970G   & 29.13(11)   & \cite{freedman01} & 77$\pm$13                 & ...       \\
1973R   & 29.86(08)   & \cite{freedman01} & 87$\pm$15                 & ...       \\
1999em  & 30.34(19)   & \cite{leonard03}  & 64$\pm$13                 & 65$\pm$12 \\
\hline
Average &             &                   & 75$\pm$7                  & 65$\pm$12 \\
\hline
\end{tabular}
\label{H0.tab}
\end{table}

HP02 found a value of $H_0$=55$\pm$12 based on one calibrator (SN~1987A),
which proves significantly lower than the current 65-75 range. The main reason
for this difference is that SN~1987A is not a plateau event and should not have been
included in the HP02 sample since the physics of its lightcurve is different
than that of SNe~IIP.

\section{Conclusions and Discussion}

This sample of 24 SNe~IIP shows that the luminosity-velocity relation 
can be used to standardize the luminosities of these objects.
The resulting Hubble diagram has a dispersion of 0.3 mag,
which implies that SNe~IIP can produce distances with a precision of 15\%.
Using four nearby SNe with Cepheid distances I find 
$H_0(V)$=75$\pm$7 and $H_0(I)$=65$\pm$12. These values compare
with $H_0$=71$\pm$2 derived from SNe~Ia \cite{freedman01},
which lends further credibility to the SCM.
 
This study confirms that SNe~IIP offer great potential as distance
indicators. The recently launched Carnegie Supernova Program at Las Campanas Observatory has
already targeted $\sim$20 such SNe and in the next years it will
produce an unprecedented database of spectroscopy and photometry
for $\sim$100 nearby SNe, which will be ideally suited for
cosmological studies.

Although the precision of the SCM is only half as good as that produced by
SNe~Ia, with the 8-m class telescopes currently in operation it should be possible
to get spectroscopy of SNe~IIP down to $V$$\sim$23 and start populating the Hubble diagram
up to $z$$\sim$0.3. A handful of SNe~IIP will allow us to get and
independent check on the distances to SNe~Ia.

\vspace{0.05in}

Support for this work was provided by NASA through Hubble Fellowship
grant HST-HF-01139.01-A awarded by the Space Telescope Science Institute,
which is operated by the Association of Universities for Research in Astronomy,
Inc., for NASA, under contract NAS 5-26555.

%
%

%

\begin{thebibliography}{99.}
%
%
%




\def\aa{{A\&A~}}
\def\aas{{A\&AS~}}
\def\aj{{AJ~}}
\def\annrev{{ARA\&A~}} 
\def\apj{{ApJ~}}
\def\apjs{{ApJS~}}
\def\baas{{BAAS~}}
\def\mnras{{MNRAS~}}
\def\nat{{Nature~}}
\def\pasp{{PASP~}} 


\bibitem{barbon79} R.~Barbon, F.~Ciatti, L.~Rosino: \aa \textbf{72}, 287 (1979)





\bibitem{eastman96} R.G.~Eastman, B.P.~Schmidt, R.~Kirshner: \apj \textbf{466}, 911 (1996)


\bibitem{freedman01} W.L.~Freedman et al: \apj \textbf{553}, 47 (2001)

\bibitem{hamuy01} M.~Hamuy: Type II supernovae as distance indicators. Ph.D. Thesis, Univ. Arizona,
Tucson (2001)

\bibitem{hamuy02} M.~Hamuy, P.A.~Pinto: \apj \textbf{566}, L63 (2002) (HP02)

\bibitem{hamuy03a} M.~Hamuy: Review on the observed and physical properties of core collapse supernovae.
In: \textit{Core Collapse of Massive Stars}, ed by C.L.~Fryer (Kluwer, Dordrecht 2003a) in press (astro-ph/0301006)

\bibitem{hamuy03b} M.~Hamuy: The latest version of the standardized candle method for type II supernovae.
In: \textit{Carnegie Observatories Astrophysics Series}, vol 2, ed by W.L.~Freedman
(Cambridge, Cambridge Univ. Press 2003b) in press (astro-ph/0301281)


\bibitem{leonard03} D.C.~Leonard, S.M.~Kanbur, C.C.~Ngeow, N.R.~Tanvir: \apj, in press (2003) (astro-ph/0305259)
 
\bibitem{litvinova83} I.Y.~Litvinova, D.K.~Nadezhin: Ap\&SS \textbf{89}, 89 (1983)

\bibitem{litvinova85} I.Y.~Litvinova, D.K.~Nadezhin: SvAL \textbf{11}, 145 (1985)

\bibitem{parodi00} B.R.~Parodi, A.~Saha, A.~Sandage, G.A.~Tammann: \apj \textbf{540}, 634 (2000)


\bibitem{schlegel98} D.J.~Schlegel, D.P.~Finkbeiner, M.~Davis: \apj \textbf{500}, 525 (1998)

\bibitem{thim03} F.~Thim et al: \apj \textbf{590}, 256 (2003)

\bibitem{tonry00} J.L.~Tonry, J.P.~Blakeslee, E.A.~Ajhar, A.~Dressler: \apj \textbf{530}, 625 (2000)


\bibitem{weiler02} K.W.~Weiler, N.~Panagia, M.J.~Montes, R.A.~Sramek: \annrev \textbf{40}, 387 (2002)



\end{thebibliography}
%

\printindex
\end{document}